\documentclass[journal, 10pt]{IEEEtran}

\usepackage{esvect}
\usepackage{yhmath}
\usepackage{epsfig}
\usepackage{times}
\usepackage{cite}
\usepackage{amsmath,amssymb}
\usepackage{subfigure}
\usepackage{setspace}

\usepackage{graphicx}
\begin{document}

\title{A Geometrical-Based Throughput Bound Analysis \\for
Device-to-Device Communications \\in Cellular Networks}
\author{
Minming~Ni,~\IEEEmembership{Member,~IEEE, } 
Lei~Zheng,~\IEEEmembership{Student Member,~IEEE,} 
~Fei~Tong, \\
Jianping~Pan,~\IEEEmembership{Senior Member,~IEEE,} 
~Lin~Cai,~\IEEEmembership{Senior Member,~IEEE} 
}

\maketitle
\thispagestyle{empty}

\begin{abstract}
Device-to-device (D2D) communications in cellular
networks are promising technologies for improving network throughput,
spectrum efficiency, and transmission delay. In this paper, we first
introduce the concept of guard distance to explore a proper system
model for enabling multiple concurrent D2D pairs in the same cell.
Considering the Signal to Interference Ratio (SIR) requirements for
both macro-cell and D2D communications, a geometrical method is
proposed to obtain the guard distances from a D2D user equipment (DUE)
to the base station (BS), to the transmitting cellular user equipment
(CUE), and to other communicating D2D pairs, respectively, when the
uplink resource is reused. By utilizing the guard distances, we then
derive the bounds of the maximum throughput improvement provided by
D2D communications in a cell. Extensive simulations are conducted to
demonstrate the impact of different parameters on the optimal maximum
throughput. We believe that the obtained results can provide useful
guidelines for the deployment of future cellular networks with
underlaying D2D communications. 
\end{abstract}

\section{Introduction} 

In recent years, device-to-device~(D2D) communications underlaying a
cellular infrastructure have received plenty of attention from both
academia and industry. With D2D communications, a user equipment~(UE)
can directly exchange data with another one in its proximity, instead
of having the base station~(BS) as the relay node. By facilitating the
reuse of cellular spectrum resources, D2D communications are promising
in reducing transmission delay, increasing cell throughput, and
enhancing spectrum efficiency. Thus, the D2D communications have been
considered as one of the key components in the next-generation
broadband cellular networks, such as the Third-Generation Partnership
Project~(3GPP) Long Term Evolution Advanced (LTE-A). Currently, the
related work items are being standardized as LTE Direct in 3GPP as a
Release 12 feature of LTE. Moreover, specific business models for
different D2D usage cases are also being studied by wireless operators
and vendors~\cite{Lei:12MCom}.

One of the serious challenges for D2D communications in cellular
networks is the interference between D2D UEs~(DUE) and other cellular
UEs~(CUE). When a pair of DUEs communicate using the uplink cellular
resources, the D2D communication might be affected by the simultaneous
transmission between a CUE and the BS. Moreover, if there are multiple
concurrent D2D pairs, the accumulated interference may also influence
the quality of the signal received by the BS. Similarly, when a D2D
communication reuses the downlink resources, a transmitting DUE might
cause reception failures of its nearby CUEs. It is still an open issue
to effectively allocate the radio resources among DUEs and CUEs in
cellular networks with underlaying D2D communications.

In this paper, we focus on an uplink resource reusing scenario, in
which one CUE and multiple D2D pairs are transmitting simultaneously.
We first investigate the \emph{guard distances} from a DUE to the BS,
to the transmitting CUE, and to other communicating D2D pairs,
respectively. The basic ideas are: 1) a DUE receiver has to stay away
with certain distances from the transmitting CUE and other DUE
transmitters; and 2) all the DUE transmitters should keep a certain
distance away from the BS, so that all the signal receptions can be
successfully achieved. By utilizing the obtained guard distances, we
then analyze the maximum number of concurrent D2D communications that
can be carried within the observed cell, which is further used to study
the influences of different parameters on the optimal throughput
improvement. 

The main contributions of this paper are twofold. First, we propose a
geometrical method to arrange the concurrent D2D pairs in a cell to
maximize the spectrum reuse, which is based on our analysis results 
of the interference-free guard distances. Second, we obtain the maximum
throughput improvement of D2D communications in a single cell as a
piecewise function of the distance between the transmitting CUE and
the BS, which could be further used as a basis to design the radio
resource allocation schemes for the UEs in a cellular network with
underlaying D2D communications. To the best of our knowledge, this
paper is one of the first to systematically study the throughput
performance with multiple concurrent D2D pairs.

The remainder of the paper is organized as follows. In
Section~\ref{sec:relwork}, we summarize the related work in radio
resource allocation and throughput analysis. The system model for our
analysis is described in Section~\ref{sec:sysMod}. In
Section~\ref{sec:ana}, the closed-form expressions of the guard
distances are obtained and followed by the derivation for the bounds
of the maximum throughput improvement in a single cell. Simulation
results are presented in Section~\ref{sec:eval}. A discussion about
the possible future work is given in Section~\ref{sec:dis}, and
Section~\ref{sec:con} finally concludes this paper.

\section{Related Work}\label{sec:relwork}

The idea of capturing the benefits of the proximity between network
nodes has been studied to improve cellular network's radio
coverage~\cite{Aggelou:01PC}, traffic balance~\cite{Wu:01JSAC}, user
fairness~\cite{Luo:03Mobicom}, and other performance metrics for quite
a long time. However, these efforts usually assume that the local data
exchanges utilize an unlicensed frequency band, such as the 2.4 GHz
Industry Science Medicine (ISM) spectrum, rather than reusing the
spectrum resources allocated for cellular networks. Given that the
quality of service (QoS) in the unlicensed spectrum may fail to be
controlled or guaranteed, the underlaying D2D communications are more
preferable to both service providers and device vendors. Currently,
there are multiple ongoing research topics in this area, including
mode selection~\cite{Min:11TWC-a}, scheduling~\cite{Han:12VTCf},
resource management~\cite{Lee:13WCNC}, etc. A detailed survey of the
design challenges and potential solutions for the D2D communications
can be found in \cite{Fodor:12MCom}. In this paper, we are interested
in the effect of interference on throughput performance of the
underlaying D2D networks. To control/coordinate the interference and
improve the throughput performance, existing work can be roughly
classified into two categories, including radio resource allocation
and the theoretical analysis of the achievable performance bounds. 

For the radio resource allocation, one of the important early work is
\cite{Janis:09OtherJ}, in which an initial framework for the D2D
communications in the cellular networks was proposed. According to its
simulation results, a D2D communication could be enabled without
degrading the performance of the cellular network, even in an
interference-limited scenario with heavy traffic load. In
\cite{Yu:11TWC}, by assuming that the radio resource managements were
adopted for both the cellular and D2D connections, three possible
resource allocation methods, i.e., non-orthogonal, orthogonal, and
cellular operation, were studied. Moreover, two optimization cases,
greedy sum-rate maximization and sum-rate maximization with rate
constraints, were also analyzed in~\cite{Yu:11TWC}. In
\cite{Chae:11APCC}, a radio resource allocation scheme was proposed
for D2D communications underlaying cellular networks with fractional
frequency. The different frequency bands utilized by DUEs and CUEs
were chosen according to whether the UEs were located in the inner or
outer region of a cell, so that the interference could be greatly
alleviated. In one of the most recent work~\cite{Erturk:2013LC}, under
the power control constraint, the spatial distribution of a D2D
network's transmission power and the signal to interference plus noise
ratio (SINR) were derived based on the homogeneous Poisson Point
Processes (PPP). In general, most of the existing resource management
schemes focused on the scenario that each cell had one D2D pair and
one CUE (or multiple ones when directional antennas were applied at
the BS) transmitting at the same time. The proper design guidelines to
support multiple concurrent D2D pairs within one cell are still
unclear for the radio resource allocation schemes.

For the theoretical analysis of the achievable performance bounds,
available results are relatively fewer. In \cite{Cheng:11OtherC}, the
uplink capacity gain was derived when one D2D link was enabled in an
FDD CDMA-based cellular cell. In \cite{Min:11TWC-b}, an
interference-limited area~(ILA) control scheme was proposed to manage
the interference from CUEs to a D2D transaction when multiple antennas
were used by the BS. By analyzing the coverage of ILA, a lower bound
of the ergodic capacity was also obtained for DUEs using uplink
cellular radio resources. After that, a similar approach was extended
to the downlink resources sharing scenario in \cite{Chen:12OtherC}. In
\cite{Liu:12CrownCom}, the maximum achievable transmission capacity,
which was defined as the spatial density of successful transmissions
per unit area, was analyzed for the hybrid D2D and cellular network
through stochastic geometry. However, due to the inevitable
interference accumulated at the BS, most of the existing analytical
results assuming a single D2D pair in a cellular network cannot be
directly extended to a scenario with multiple coexisting D2D pairs.
Therefore, the performance of D2D communications in the latter is
still an under-developed issue, which could further improve the
spectrum efficiency and increase the cellular network throughput.

To make up the shortage of performance bounds analysis for D2D
communications in cellular networks, some useful insights might be
obtained from the existing results of the Protocol Interference Model
(PrIM) and the Physical Interference Model (PhIM)-based capacity
analyses, which were mainly initialized from \cite{Gupta:2000ht}. By
introducing a spatial protection margin $\Delta$, PrIM defines a
location-based condition for successful communications between a
single node pair. The condition could be applied to all the concurrent
node pairs in the network to obtain the capacity bounds for different
network settings, for example, the effect of directional antennas on
network capacity bounds were studied in~\cite{Yi:2003wh}. However,
PrIM does not take into account the aggregated interference, which
happens to be vital for the D2D scenario, e.g., the constraint on the
total interference power accumulated at the BS. Comparing with PrIM,
PhIM is based on the power capture model, and focuses on the
aggregated interference on a specific receiver. By assuming the
interference power follows a Gaussian distribution, the General PhIM
was proposed in \cite{Agarwal:2004vz}. Moreover, a series of
graph-based interference models have also been developed based on PhIM
for different research purposes and network scenarios
\cite{Brar:2006uu,Kim:2007fv}. However, the higher computation
complexity, which is caused by calculating the sum of all the
undesired signals, might also prevent the application of PhIM on large
and complicated network scenarios, e.g., the D2D communications
deployed in an irregular network area.

\section{System Model}\label{sec:sysMod}

In this paper, we aim to study the network throughput improvement when
multiple D2D communications coexist with a single CUE-BS
communication\footnote{This one CUE model presents the situation that
orthogonal channel resources are allocated to different CUEs in
traditional cellular networks such as GSM. The more general multiple
CUEs scenario will be one of our future work items, which will be
briefly discussed in Section~\ref{sec:dis}.}. To develop a tractable
model of D2D communications in a single cell, it is assumed that the
coverage area of a BS is a disk with radius $r_\mathrm{C}$. Two
working modes, CUE and DUE, are available for each UE. In the CUE
mode, a UE sends packets via the BS; while UEs in the DUE mode
exchange packets via direct connections in an ad hoc style utilizing
the uplink radio resource~\cite{Doppler:09MCom}. The case when the
downlink radio resource is reused, which can still be analyzed by the
method developed in this paper, will be one of our research issues in
the near future.

To describe a signal's power attenuation, a general path-loss 
model is applied here as
\begin{equation}
\label{eq:path-loss} P_{\mathrm r} = \frac{\beta}{d^{\,\alpha}}
P_\mathrm{t} = L(d) P_{\mathrm t}\quad, 
\end{equation} 
where $P_\mathrm{t}$ is the transmission power, $P_\mathrm{r}$ is the
average received signal power at distance $d$ from the transmitter,
$\beta$ is the path-loss constant determined by the hardware features
of the transceivers, and $\alpha$ is the path-loss exponent depending
on the propagation environment \cite{Rappaport:01Book}. For better
readability, we use $L(d)$ to represent the path-loss ratio of the
transmission power at distance $d$. Moreover, $L_\mathrm{B}(d)$ and
$L_\mathrm{D}(d)$ are used to represent the different physical
characteristics and constraints of CUE-BS and DUE-DUE links,
respectively. This model can be extended to study the instantaneous
throughput or throughput distribution by introducing a lognormal
random variable representing the channel shadowing effect, which will
be another topic of our future research.

Since a CUE's transmission can always be coordinated by its BS, power
control schemes could be implemented to achieve different design
goals. For compensating the near-far effect , we assume that the average
received signal power at a BS is controlled to the same level,
$P_{\mathrm{r, CB}}$, for each CUE~\cite{Min:11TWC-b}. Therefore, the
maximum CUE transmission power $P_{\mathrm{t, C_{max}}}$ is utilized
when the CUE is located at the boundary of the cell, and $P_\mathrm{r,
CB} = L_\mathrm{B} (r_\mathrm{C})\, P_\mathrm{t, C_{max}}$.

Compared with the CUE-BS connection, the coordinations among D2D
communications are usually limited, and are more vulnerable to the
unexpected channel conditions. Thus, we assume that all the D2D
communications are carried with a fixed transmission power
$P_{\mathrm{t, D}}$ and a constant bit rate $R_b$. Moreover, we define
that a D2D connection will only be established when the distance
$d_{\mathrm{D2D}}$ between the two DUEs is within a predefined range
$[d_{\min}, d_{\max}]$, where $d_{\min}$ is a very short distance
representing the minimum physical separation between any two UEs.

\begin{figure}[!h] \centering
\includegraphics[scale = 0.75]{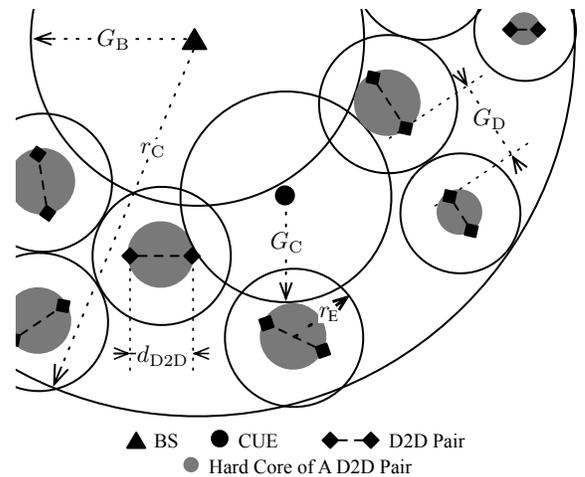} 
\caption{An illustration of the DUE-CUE coexisting scenario} 
\label{fig:model3r} 
\end{figure}

With multiple D2D pairs reusing the uplink resource in a cell,
interference occurs between DUE and DUE, CUE and DUE, DUE and
BS\footnote{Currently, the interference generated by the DUEs in the
nearby cells is not considered. This approximation is reasonable as
long as the frequency reuse factor is larger than one, which means
neighboring cells are allocated with different uplink resources, so
this kind of interference can be just ignored.}. For successful
receptions, we assume that two predefined Signal to Interference Ratio
(SIR) thresholds should be achieved at a DUE as~$\delta_\mathrm{D}$,
and at the BS as $\delta_\mathrm{B}$, respectively. 
To guarantee $\delta_\mathrm{D}$, a DUE receiver should stay at a
guard distance $G_\mathrm{D}$ away from all the other transmitting
DUEs, and at a guard distance $G_\mathrm{C}$ away from the
transmitting CUE to limit the received total interference. Similarly,
to satisfy the required receiving SIR $\delta_\mathrm{B}$ at the BS,
there should also be a minimum guard distance $G_\mathrm{B}$ between
the BS and all the DUE transmitters, which limits the number of
concurrent D2D pairs in a cell and the total interference accumulated
at BS. In addition, considering that D2D communications are usually
bidirectional (e.g., service discovery, data transmission, and ACK
feedback) and the distance between a D2D pair is typically short, the
transmitter and receiver of \text{a D2D pair} will not be
distinguished in our interference analysis\footnote{This assumption
could be verified by simulation, and the related results will be shown
in Fig.~\ref{fig:sim5}, Section~\ref{sec:simu-c}.}. Therefore, the
\emph{Exclusive Region} (ER) occupied by a D2D pair with link distance
$d_\mathrm{D2D}$ can be modeled as a disk with radius $r_\mathrm{E} =
(d_\mathrm{D2D} +G_\mathrm{D})/2$. Due to the possible difference of
$d_\mathrm{D2D}$ for each communicating pair, $r_\mathrm{E}$ could be
different for each ER. Moreover, we define $r_\mathrm{E, min} =
(G_\mathrm{D} + d_\mathrm{min})/2$ and $r_\mathrm{E, max} =
(G_\mathrm{D} + d_\mathrm{max})/2$. For concurrent transmissions, two
D2D pairs' ERs should not overlap with each other. 
The three guard distances and ERs are depicted in
Fig.~\ref{fig:model3r}, which demonstrates a part of a cell with the
coexistence of of one CUE and several D2D pairs. For an ER, we use a
grey disk to illustrate an imaginary \emph{hard core}, whose diameter
is the distance between the D2D pair. Since the boundary of a hard
core represent all the possible relative positions of the D2D pair as
long as $d_\mathrm{D2D}$ is fixed, as shown in the figure, neither the
BS guard region, which is a disk centered at the BS with radius
$G_\mathrm{B}$, nor the CUE's impact disk, which is a disk centered at
the CUE with radius $G_\mathrm{C}$, could intersect with any D2D
pair's hard core. Moreover, all the hard cores have to stay inside the
cell's coverage area. 

\section{Theoretical Analysis}\label{sec:ana}

\subsection{Parameter Setting for Guard Distances}
Before investigating the bounds of the throughput improvement brought
by the D2D communications in a single cell, the three guard distances
$G_\mathrm{B}$, $G_\mathrm{C}$, and $G_\mathrm{D}$ will be determined
for a reasonable setting first in this subsection. 

\subsubsection{Calculations for $G_\mathrm{D}$}
\begin{figure}[!h]
\centering
\includegraphics[scale = 0.7]{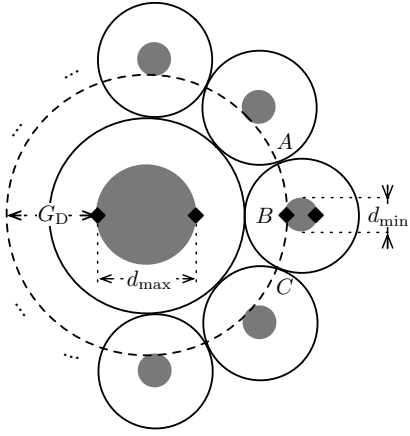} 
\caption{An illustration for calculating $G_\mathrm{D}$} \label{fig:GD}
\end{figure}

When no CUE is considered, for a DUE node, the worst interfered
scenario happens when: a) its link distance reaches the maximum
allowable value $d_{\max}$; and b) it is affected by the maximum
possible number of nearby D2D pairs. Owing to that $G_\mathrm{D}$ is a
fixed parameter for all the D2D pairs, the maximum number of ERs
surrounding the observed D2D pair is obtained when $d_\mathrm{D2D} =
d_{\min}$ holds for all the other D2D pairs, as shown in Fig.
\ref{fig:GD}. With acceptable accuracy, only the surrounding D2D pairs
in the first layer, which generate the majority part of the total
interference, are considered here. 

The number of the first layer surrounding D2D pairs could be calculated as
\begin{equation}
n_\mathrm{s} = \left\lfloor 
\frac{2\pi (r_\mathrm{E, max} + \frac{G_\mathrm{D}}{2})}{\mathcal{L} 
\left(r_\mathrm{E, max} + \frac{G_\mathrm{D}}{2},~
 r_\mathrm{E, min},~ r_\mathrm{E, min} + r_\mathrm{E, max}\right)} 
\right\rfloor~~,
\end{equation}
where the function $\mathcal{L}$ in the denominator is used to
calculate the length of the arcs $\wideparen{ABC}$ in Fig.
\ref{fig:GD}. The detailed expression of function $\mathcal{L}$ is
given in the Appendix. Since $G_\mathrm{D}$ is the minimum distance
between the interfered DUE and all the interferers, the transmission
bit rate $R_b$ should be at least higher than the bit rate obtained in
the worst interfered scenario, in which all the $n_\mathrm{s}$
surrounding transmitters are located with distance $G_\mathrm{D}$ to
the observed DUE, as
\begin{equation}
R_b > W \log \left( 1+ \frac{P_\mathrm{t,D} L _\mathrm{D}(d_{\max}) }
{N_0 W + n_\mathrm{s} \cdot P_\mathrm{t, D} L_\mathrm{D}(G_\mathrm{D})} 
\right)\quad,
\end{equation}
where $W$ is the system bandwidth, and $N_0$ is the one-sided spectral
density of additive white Gaussian noise (AWGN). Finally, $G_\text{D}$
could be calculated by combining with the path-loss model in
(\ref{eq:path-loss}) as
\begin{equation}
 G_\mathrm{D} =  d_{\max} \sqrt[\alpha'] 
 {\frac{\beta' \, n_\mathrm{s}\, P_{\mathrm{t,D}}\,(2^{R_b/W} -1)}
 {\beta' P_{\mathrm{t,D}} - N_0 W \,d_{\max}^{\,\alpha'}  (2^{R_b/W} 
 -1)}}\quad,
\end{equation}
where $\alpha'$ and $\beta'$ are the path-loss parameters associated with 
$L_\mathrm{D}(d)$.

\subsubsection{Calculations for $G_\mathrm{C}$}
Based on the system model, no matter how many D2D pairs can be
activated simultaneously, the effects of a CUE transmission's
interference on each communicating DUE node are independent of each
other. Suppose a CUE node is located with distance $d_{\mathrm{CB}}$
to its BS, then its transmission power $P_\mathrm{t,C}$ can be
represented as 
\begin{equation}
	P_\mathrm{t,C} = \frac{P_\mathrm{r,CB}}{L_\mathrm{B}(d_\mathrm{CB}) } 
	  = P_{\mathrm{t,C_{\max}}}  \cdot \frac{L_\mathrm{B}(r_\mathrm{C})}
	  {L_\mathrm{B}(d_\mathrm{CB})} \quad.
\end{equation}
Similarly, considering the SIR constraint at DUE, we have
\begin{equation}
\delta_\mathrm{D} \leq \frac{L_\mathrm{D} (d_\mathrm{D2D}) P_\mathrm{t,D} }
{ L_\mathrm{D}(d_\mathrm{CD}) P_\mathrm{t,C} } = 
\frac{L_\mathrm{D}(d_\mathrm{D2D}) \, L_\mathrm{B}(d_\mathrm{CB}) \cdot P_\mathrm{t,D} }
{L_\mathrm{D}(d_\mathrm{CD}) \, L_\mathrm{B}(r_\mathrm{C}) \cdot P_{\mathrm{t, C_{\max}}} } \quad,
\end{equation}
where $d_\mathrm{CD}$ is the distance between the CUE and the
interfered DUE node. For a CUE, the worst case is that the being
affected D2D pair has the longest link distance ($d_\mathrm{D2D} =
d_{\max}$), which should be used to determine $G_\mathrm{C}$ for
system setting. By combining the path-loss model in
(\ref{eq:path-loss}) and ignoring the small difference between the
path-loss exponent of the CUE-BS and DUE-DUE links \cite{Min:11TWC-b},
$G_\mathrm{C}$ could be obtained as 
\begin{equation} \label{eq:R2}
G_\mathrm{C} = K\cdot d_{\mathrm{CB}} \quad,
\end{equation}
where $K$ is a function of the DUE transmission power $P_\mathrm{t,D}$ as 
\begin{equation}\label{eq:K}
K = \frac{d_{\max} \sqrt[\alpha]{\delta_{\mathrm{D}} P_{\mathrm{t, C_{\max}} }}}
{ r_{\mathrm{C}}} \cdot \frac{1}{\sqrt[\alpha]{P_\mathrm{t,D}  } }\quad. 
\end{equation}

\subsubsection{Calculations for $G_\mathrm{B}$}
For an observed BS, the worst interfered case happens when the number
of D2D paris in its cell reaches the theoretical upper bound, which
means: a) the CUE's impact disk is fully included in the BS guard
region, so it has no negative effect on any D2D communication; and b)
the condition $d_\mathrm{D2D} = d_{\min}$ holds for each D2D pair, so
each ER only occupies the smallest area. 
However, accurately calculating the maximum number of disks that
could be arranged into a given ring area without intersection is still
an open issue currently, which is known as a case of the 
\emph{circle/sphere
packing} problem in geometry \cite{Stephenson:03circlepacking}. 
\begin{figure}[!h]
\centering
\includegraphics[scale = 0.7]{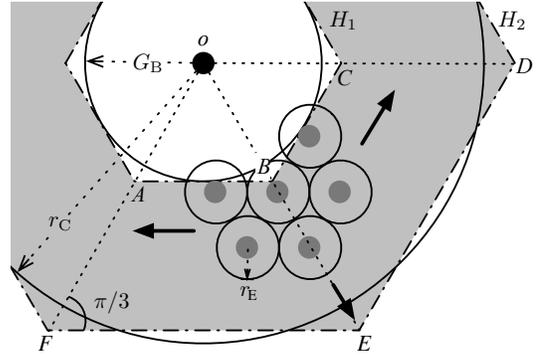} 
\caption{Hexagon approximation for the observed cell} \label{fig:hex-appox}
\end{figure}

To solve the problem, two hexagons $H_1$ and $H_2$ are used to
approximate the BS's guard disk and the cell coverage area,
respectively, as shown in Fig. \ref{fig:hex-appox}. Then the ring area
is transferred as the grey region consisting of six identical
isosceles trapezoids with $\pi/3$ base angles. The side length of
hexagon $H_1$ is set to $r_{\mathrm{H}_1} = \overline{oB} =
2G_\mathrm{B}/\sqrt{3}$, which means that the hexagon is circumscribed
with the guard disk of BS, so the SIR requirement $\delta_\mathrm{B}$
can still be achieved. For the hexagon $H_2$, we let the six
trapezoids' total area in Fig.~\ref{fig:hex-appox} equal the ring area
determined by $G_\mathrm{B}$ and $r_\mathrm{C}$, so its side length
$r_{\mathrm{H}_2} = \overline{oE}$ has to satisfy: 
\begin{equation}
\pi \left(r_\mathrm{C}^2 - G_\mathrm{B}^2\right) 
= \frac{3\sqrt{3}}{2}\left(r_{\mathrm{H_2}}^2 - r_{\mathrm{H_1}}^2\right)\quad.
\end{equation}
Therefore, 
\begin{equation}
r_{\mathrm{H}_2} = \frac{\sqrt{2}}{3} \sqrt{\sqrt{3}\pi r_\mathrm{C}^2  - 
\left(\sqrt{3} \pi - 6\right)G_\mathrm{B}^2 } 
\quad.
\end{equation}

Due to  the symmetry feature, we can focus on one third of the approximated ring 
area initially, e.g., the polygon region $ABCDEF$ in Fig. \ref{fig:hex-appox}. 
According to the system model, the most compact way to arrange D2D pairs 
is to locate the first ER's center on line segment $BE$, while node $B$ 
is on the boundary of the ER's hard core, and arrange other D2D ERs 
without overlapping along three directions $\vv{BA}$, $\vv{BE}$, and $\vv{BC}$, 
as shown in the 
figure\footnote{The arrangement could also be made along directions
$\vv{BA}$, $\vv{EB}$, and $\vv{BC}$ (from the cell boundary to the
inside). For a network area large enough, these two arrangements lead
to almost identical results, so we only consider the previous
arranging method in this paper.}. 
Note that, this arrangement is identical to the 
\emph{hexagon packing}, which is the densest packing possible on a flat 
surface \cite{Stephenson:03circlepacking}. In this way, the maximum layers 
of D2D ERs that could be arranged in a trapezoid area can be calculated as 
\begin{equation}\label{eq:nr}
n_\mathrm{L}= \left\lfloor   
\frac{r_\mathrm{H_2} - r_\mathrm{H_1} - d_{\min}}{2r_\mathrm{E, min}}  
\right\rfloor +1 \quad.
\end{equation}
According to the geometry, in the observed polygon region $ABCDEF$, the 
number of DUE ERs that could be placed in the $i$-th layer can be calculated as
\begin{equation}
	n_i = n_i' + n_i''\quad, \label{eq:nd}
\end{equation}
where $i \in [1, n_\mathrm{L}]$,  
\begin{equation}
	n_i' = \left\lfloor   \frac{r_\mathrm{H_1} + d_{\min}/2 + (2i-3)r_\mathrm{E, min}}
	{2r_\mathrm{E, min}}\right\rfloor\quad,
\end{equation}
represents the number of the ERs in the $i$-th layer that could be
arranged within the trapezoid $ABEF$, excluding the ones on the
boundary $BE$, and
\begin{equation}
	n_i'' = \left\lfloor   \frac{r_\mathrm{H_1} + d_{\min}/2 + 2(i-1)r_\mathrm{E, min}}
	{2r_\mathrm{E, min}}\right\rfloor +1\quad,
\end{equation}
represents the number of the ERs in the $i$-th layer that could be
arranged within the trapezoid $BCDE$, including the ones on the
boundary $BE$. Note that, when ERs are arranged near the two side
boundaries of the observed polygon region ($CD$ and $AF$), if more
than half an ER's hard core could be arranged within the boundary, the
ER should be counted on one side only. Then the double counting errors
will not happen. 
Therefore, the maximum number of ERs that could be arranged in the ring 
area determined by $G_\mathrm{B}$ and $r_\mathrm{C}$ can be approximated 
as follows
\begin{equation}\label{eq:nmax}
n_{\max} \approx 3  \sum_{i =1}^{n_\mathrm{L}} n_i = \widehat{n}_{{\max}} \quad.
\end{equation}

Once $ \widehat{n}_{{\max}}$ is obtained, the total interference generated 
by D2D communications accumulating at the BS can be calculated as
\begin{equation}\label{eq:I}
I \approx 3 \sum_{i=1}^{n_\mathrm{L}} \left( 
\sum_{j=1}^{n_i'} L_\mathrm{B}(d_{i,j}) P_\mathrm{t,D}   
+ \sum_{k=1}^{n_i''} L_\mathrm{B}(d_{i,k}) P_\mathrm{t,D} \right)\quad, 
\end{equation}
where $d_{i,j}$ and $d_{i,k}$ represent the corresponding distance between 
BS and the center of the $j$-th ER in the $i$-th layer within $ABEF$
, and the center of the $k$-th ER in the $i$-th layer within $BCDE$, 
respectively, which are used to approximate the distance between BS 
and transmitting DUEs. Because $r_\mathrm{C} > G_{\mathrm{B}} \gg d_{\min}$, 
this approximation is acceptable. $d_{i,j}$ and $d_{i,k}$ could be calculated 
according to the Law of Cosine as follows
\setlength{\arraycolsep}{0.14em}
\begin{eqnarray}
d_{i,j} &=& \sqrt{ \kappa_i^2 + \kappa_j^2 - \kappa_i \kappa_j }\quad, \label{eq:dij}\\
d_{i,k} &=& \sqrt{ \kappa_i^2 + \kappa_k^2 - \kappa_i \kappa_k }\quad,\label{eq:dik}
\end{eqnarray}
where 
$\kappa_i = r_\mathrm{H_1} + d_{\min}/2 + 2(i-1)r_\mathrm{E, min}$, 
$\kappa_j = 2 j r_\mathrm{E, min}$, and
$\kappa_k = 2(k-1)r_\mathrm{E, min}$.

Due to the SIR requirement, we also have
\begin{equation}\label{eq:sirB}
\delta_\mathrm{B} \leq P_\mathrm{r, CB} / I \quad.
\end{equation}
By combining equations (\ref{eq:nr})--(\ref{eq:sirB}) and the definition 
of $r_{\mathrm{H}_1}$, $G_\mathrm{B}$ can be obtained in a numerical way 
(e.g., by the bisection search algorithm). Another byproduct here is that, 
when the CUE's impact disk is fully covered by the BS guard region, the 
upper bound of the maximum throughput improvement could be calculated as 
\begin{equation}
\mathcal{T}_\mathrm{U} =  \widehat{n}_\mathrm{max} R_b\quad. \label{eq:upper1}
\end{equation}

\subsection{Bounds for Throughput Improvement in General Scenarios}
\label{sec:general bounds}

\begin{figure}[!htb]
\centering 
\includegraphics[scale = 0.8]{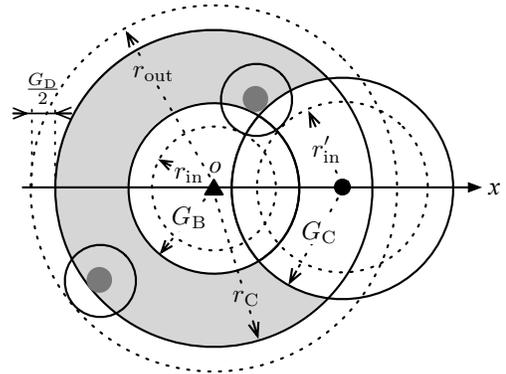} 
\caption{A general case for deploying D2D pairs} \label{fig:general} 
\end{figure}

It is clear that the actual capacity improvement is determined by the
number of concurrent transmitting D2D pairs in the network. In a
general scenario, the transmitting CUE's impact disk could move out
the guard region of the BS, so the total area for deploying
non-overlapped D2D ERs is reduced, which may further affect the
maximum throughput improvement in the cell. Due to the circular
symmetry, a Cartesian coordinate system can be built with its origin
at the observed BS, and the CUE's impact disk can always be aligned
with the $x$-axis as shown in Fig.~\ref{fig:general}, where
$d_\mathrm{CB} \in (0, r_\mathrm{C}]$. 
According to the system model, the hard cores cannot intersect with
neither the BS guard region nor cross the cell boundary, therefore,
the actual area for deploying possible D2D ERs is a slightly larger
ring region with inner radius $r_\mathrm{in} = G_\mathrm{B} -
G_\mathrm{D}/{2}$ and outer radius $r _\mathrm{out} = r_\mathrm{C} +
G_\mathrm{D}/{2}$.

Recall that $G_\mathrm{C} = K d_\mathrm{CB} $, so it is possible that
the CUE's impact disk might cross both the boundaries of BS guard
region and the cell coverage area, which is called
\emph{double-crossing} here. 
When $K$ is relatively small, the length of the CUE's impact disk's
diameter may still be shorter than $r_\mathrm{C} - G_\mathrm{B}$ when
the impact disk is just to move out the cell's coverage area, which
means the double-crossing never happens. Therefore, the range of $K$
for no double-crossing could be written as $(0, K_\mathrm{th1}]$, where
$K_\mathrm{th1}$ could be derived by the critical condition mentioned
above as 
\begin{equation}
K_\mathrm{th1} = \frac{r_\mathrm{C} - G_\mathrm{B}}{r_\mathrm{C} + G_\mathrm{B}}\quad.
\end{equation}
When $K > K_\mathrm{th1}$, if $G_\mathrm{C}$ is still shorter than 
$r_\mathrm{C} - G_\mathrm{B}$ given the CUE is on the boundary of the BS's 
guard region, the double-crossing will only happen for a specific range of 
$d_\mathrm{CB}$, and the CUE's impact disk will not intersect with the BS's 
guard region anymore. Therefore, the second threshold for $K$ can be obtained as
\begin{equation}
K_\mathrm{th2} = \frac{r_\mathrm{C} - G_\mathrm{B}}{ G_\mathrm{B}}\quad.
\end{equation}
Finally, when $K > K_\mathrm{th2}$,  once the double-crossing happens, it 
will last until $d_\mathrm{CB}= r_\mathrm{C}$. Based on these analyses, the 
area for deploying D2D ERs could be obtained as a piece-wise function 
$\mathcal{S}_{\mathrm{D}}\left(d_\mathrm{CB}, K\right)$ shown as below. 

\noindent 1) When $K \in \left(0,  K_\mathrm{th1}\right]$ :
\begin{itemize}
\item If $d_\mathrm{CB} \in [0, \frac{G_\mathrm{B}}{1+K}]$, the area for 
deploying the D2D ERs can be calculated as 
\begin{equation}
\mathcal{S}_{\mathrm{D}}\left(d_\mathrm{CB}, K\right) = \pi r _\mathrm{out}^2 
- \pi r_\mathrm{in}^2 = \mathcal{S}_{\mathrm{R}}\quad.
\end{equation}
\item  If $d_\mathrm{CB} \in (\frac{G_\mathrm{B}}{1+K}, \frac{G_\mathrm{B}}{1-K}]$, 
the CUE's impact disk crosses the boundary of BS's guard region, 
\begin{equation}
\mathcal{S}_{\mathrm{D}}\left(d_\mathrm{CB}, K\right) =  \mathcal{S}_{\mathrm{R}} 
- \pi r_\mathrm{in}'^{\,2} + \mathcal{F}\left(r_\mathrm{in}, r_\mathrm{in}', 
d_\mathrm{CB}\right)\quad,
\end{equation}
where $r_\mathrm{in}' = Kd_\mathrm{CB} - G_\mathrm{D}/2$ as shown in Fig. 
\ref{fig:general}, and the function $\mathcal{F}$ given in the Appendix is 
used to calculate the area of two disks' overlapping region.
\item If $d_\mathrm{CB} \in ( \frac{G_\mathrm{B}}{1-K}, \frac{r_\mathrm{C}}{1+K}]$, 
the CUE's impact disk is fully included in the ring area, therefore, 
\begin{equation}
\mathcal{S}_{\mathrm{D}}\left(d_\mathrm{CB}, K\right) =  
\mathcal{S}_{\mathrm{R}} - \pi r_\mathrm{in}'^{\,2} \quad.
\end{equation}
\item  If $d_\mathrm{CB} \in (\frac{r_\mathrm{C}}{1+K}, r_\mathrm{c}]$, 
part of the CUE's impact disk moves out of the cell area, therefore,  
\begin{equation}
\mathcal{S}_{\mathrm{D}}\left(d_\mathrm{CB}, K\right) =  
\mathcal{S}_{\mathrm{R}} -  \mathcal{F}\left(r_\mathrm{out}, r_\mathrm{in}', 
d_\mathrm{CB}\right)\quad. 
\end{equation} 
\end{itemize}

\noindent 2) When $K \in \left( K_\mathrm{th1},  K_\mathrm{th2}\right]$ :
\begin{itemize}
\item If $d_\mathrm{CB} \in [0, \frac{G_\mathrm{B}}{1+K}]$,  
$\mathcal{S}_{\mathrm{D}}\left(d_\mathrm{CB}, K\right) = \mathcal{S}_{\mathrm{R}}$~.
\item If $d_\mathrm{CB} \in [\frac{G_\mathrm{B}}{1+K}, \frac{r_\mathrm{C}}{1+K})$, 
the CUE's impact disk overlaps with the BS's guard region, but the 
double-crossing does not happen, 
\setlength{\arraycolsep}{0.0em}
\begin{equation}
\mathcal{S}_{\mathrm{D}}\left(d_\mathrm{CB}, K\right)  =  
\mathcal{S}_{\mathrm{R}} - \pi r_\mathrm{in}'^{\,2} + \mathcal{F}\left(r_\mathrm{in}, 
r_\mathrm{in}', d_\mathrm{CB}\right)\quad.
\end{equation}
\setlength{\arraycolsep}{5pt}
\item If $d_\mathrm{CB} \in [\frac{r_\mathrm{C}}{1+K},  
\frac{G_\mathrm{B}}{1-K})$, double-crossing happens,
\setlength{\arraycolsep}{0.0em}
\begin{eqnarray}
\mathcal{S}_{\mathrm{D}}\left(d_\mathrm{CB}, K\right) &{}={}&  
\mathcal{S}_{\mathrm{R}} + \mathcal{F}\left(r_\mathrm{in}, r_\mathrm{in}', 
d_\mathrm{CB}\right) \nonumber\\
&&{-}\:  \mathcal{F}\left(r_\mathrm{out}, r_\mathrm{in}', d_\mathrm{CB}
\right)\quad.
\end{eqnarray}
\setlength{\arraycolsep}{5pt}
\item  If $d_\mathrm{CB} \in [\frac{G_\mathrm{B}}{1-K}, r_\mathrm{C}]$, 
the CUE's impact disk only overlaps with the ring area, therefore,
\begin{equation}
\mathcal{S}_{\mathrm{D}}\left(d_\mathrm{CB}, K\right)  =  
\mathcal{S}_{\mathrm{R}}  - \mathcal{F}\left(r_\mathrm{out}, r_\mathrm{in}', 
d_\mathrm{CB}\right)\quad.
\end{equation}
\end{itemize}

\noindent 3) When $K > K_\mathrm{th2} $ :
\begin{itemize}
\item If $d_\mathrm{CB} \in [0, \frac{G_\mathrm{B}}{1+K}]$, 
$\mathcal{S}_{\mathrm{D}}\left(d_\mathrm{CB}, K\right) = \mathcal{S}_{\mathrm{R}}$~.
\item If $d_\mathrm{CB} \in [\frac{G_\mathrm{B}}{1+K}, \frac{r_\mathrm{C}}{1+K})$, 
the CUE's impact disk only overlaps the BS's guard region and the double-crossing 
does not happen, therefore,  
\setlength{\arraycolsep}{0.0em}
\begin{equation}
\mathcal{S}_{\mathrm{D}}\left(d_\mathrm{CB}, K\right) {}={}  
\mathcal{S}_{\mathrm{R}} - \pi r_\mathrm{in}'^{\,2} + \mathcal{F}
\left(r_\mathrm{in}, r_\mathrm{in}', d_\mathrm{CB}\right)\quad.
\end{equation}
\setlength{\arraycolsep}{5pt}
\item If $d_\mathrm{CB} \in [\frac{r_\mathrm{C}}{1+K},  r_\mathrm{C})$, 
double-crossing happens,
\setlength{\arraycolsep}{0.0em}
\begin{eqnarray}
\mathcal{S}_{\mathrm{D}}\left(d_\mathrm{CB}, K\right)  &{}={}&  
\mathcal{S}_{\mathrm{R}} + \mathcal{F}\left(r_\mathrm{in}, r_\mathrm{in}', 
d_\mathrm{CB}\right) \nonumber\\
&&{-}\:  \mathcal{F}\left(r_\mathrm{out}, r_\mathrm{in}', d_\mathrm{CB}
\right)\quad.
\end{eqnarray}
\setlength{\arraycolsep}{5pt}
\end{itemize}

As mentioned earlier, when all the ERs have identical radius
$r_\mathrm{E}$, the most compact arrangement of all the ERs is the
hexagon packing, in which each ER occupies a hexagon region with area
$2\sqrt{3}r_\mathrm{E}^2$. Therefore, the maximum number of concurrent
D2D pairs that could be arranged within a single cell, when $K$ and
$d_\mathrm{CB}$ are given, can be approximated by
\begin{equation}
\widehat{n}_{\max}\left(d_\mathrm{CB}, K\right) = 
\frac{\mathcal{S}_{\mathrm{D}}\!\left(d_\mathrm{CB}, K\right)}
{2 \sqrt{3} r_\mathrm{E}^2}\quad. \label{eq:nmax-conti}
\end{equation}
Note that $r_\mathrm{E}$ has two extreme situations 
($r_\mathrm{E, min}$ and $r_\mathrm{E, max}$ as defined earlier), 
which are corresponding to the minimum and maximum area an ER could cover, 
respectively. Therefore, for the general scenario, we could obtain the 
upper and lower bounds of the maximum throughput improvement in a cell as
\begin{equation}
\mathcal{T}_\mathrm{U}\left( d_\mathrm{CB}, K \right) = 
\frac{ 2 R_b \,\mathcal{S}_{\mathrm{D}}\!\left(d_\mathrm{CB}, K\right)}
{ \sqrt{3} (G_\mathrm{D} + d_{\min})^2}\quad,\label{eq:upper2}
\end{equation}
and
\begin{equation}
\mathcal{T}_\mathrm{L} \left(d_\mathrm{CB}, K\right) = 
\frac{ 2 R_b\, \mathcal{S}_{\mathrm{D}}\!\left(d_\mathrm{CB}, K\right)}
{\sqrt{3} (G_\mathrm{D} + d_{\max})^2}\quad. \label{eq:lower}
\end{equation}

\section{Performance Evaluation and Discussion}\label{sec:eval}

All our simulation results are obtained using MATLAB. While all the
analytical results can be calculated directly, a simulator is also
developed to investigate the expectations of the maximum throughput
improvement when $d_\mathrm{D2D}$ is changed to a random variable. The
common parameters for the simulation are set as: cell radius
$r_\mathrm{C} = 500$ m, minimum D2D communication range $d_{\min} = 2$
m, maximum D2D communication range $d_{\max} = 150$ m, system
bandwidth $W = 5$ MHz, and the one-sided spectral density of AWGN
power $N_0 = -174$ dBm/Hz to represent a cell in the urban scenario.
By referring to \cite{3GPP}, the path-loss ratios are set as
$L_\mathrm{B}(d) =-128.1- 37.6\lg(d/1000)$ (dB) for the BS-CUE link,
and $L_\mathrm{D}(d) =-38-37.6 \lg(d)$ (dB) for the DUE-DUE link, so
the path-loss exponent $\alpha=3.76$. In the following part of this
section, the simulation results are demonstrated in groups to show the
effect of different parameters on the throughput improvement by D2D
communications.

\subsection{Impact of $P_\mathrm{t, C_{max}}$ on 
$\mathcal{T}_\mathrm{U}$}\label{sec:Simu-1}

Intuitively, increasing DUE's transmission power $P_\mathrm{t, D}$
could support higher bit rate $R_b$ with the same guard distance
$G_\mathrm{D}$. However, the superposed interference at BS generated
by DUEs also increases with $P_\mathrm{t, D}$. As a result, the guard
distance $G_\mathrm{B}$ is enlarged and the maximum number of
concurrent D2D pairs in the observed cell is reduced. Therefore, for a
given system setting, there should be an optimal range of
$P_\mathrm{t, D}$ to obtain a relatively good performance improvement.

\begin{figure}[!htb]
\centering 
\includegraphics[scale = 0.55]{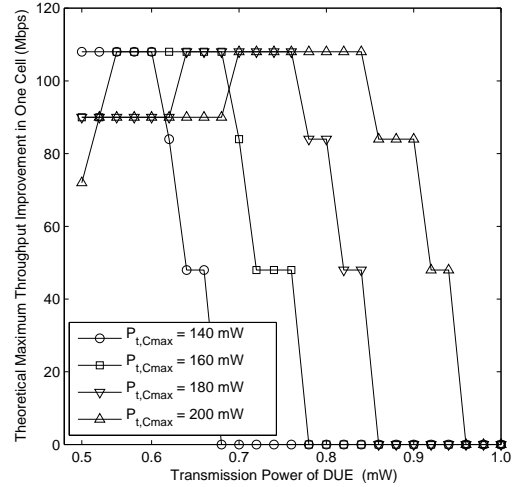} 
\caption{$P_\mathrm{t,D}$ vs. $\mathcal{T}_\mathrm{U}$ 
(when CUE is not considered) with changed $P_\mathrm{t,C_{max} }$ 
($R_b =2$ Mbps) }\label{fig:simu1} 
\end{figure}

\begin{figure}[!htb]
\centering 
\includegraphics[scale = 0.8]{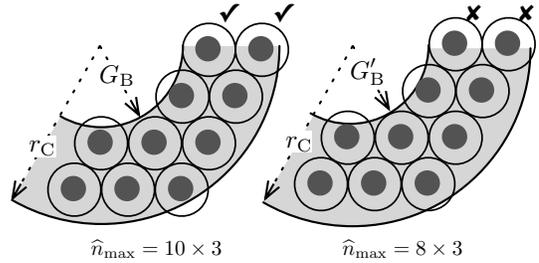} 
\caption{A demonstration for the decrease of  $\mathcal{T}_\mathrm{U}$, 
when $P_\mathrm{t,C_{max}}$ is increased}\label{fig:simu1-ex} 
\end{figure}

When the CUE's impact disk is within the guard region of the BS, the
upper bound for the maximum throughput of D2D communication
$\mathcal{T}_\mathrm{U}$ could be obtained from (\ref{eq:upper1}), and
the changing patterns of $\mathcal{T}_\mathrm{U}$ with different
$P_\mathrm{t,D}$ and $P_\mathrm{t,C_{max}}$ are shown in Fig.
\ref{fig:simu1}. As demonstrated in this figure, when $P_\mathrm{t,D}$
is relatively low, higher $P_\mathrm{t,C_{max}}$ may lead to a lower
maximum throughput improvement. This special phenomenon can be
reproduced as shown in Fig. \ref{fig:simu1-ex}, which illustrates one
third of the ring area determined by the same $r_\mathrm{C}$ but
slightly different BS guard distances, and explained as follow. When
$P_\mathrm{t, D}$ is fixed, the rise of $P_\mathrm{t,C_{\max}}$ will
contribute to the increase of the total tolerable interference signal
power at BS, which is represented as the decline of BS's guard
distance $G_\mathrm{B}$. However, since all the D2D pairs' ERs have to
be placed without overlapping while all the hard cores have to stay
within the ring area, decreasing $G_\mathrm{B}$ means fewer ERs could
be arranged in the first inner layer of the ring area. Moreover, the
increased area is accumulated at the outer ring, but the area will not
be able to be utilized until one extra ER could be located in.
Therefore, increasing $P_\mathrm{t, C_{\max}}$ may not directly result
in the rise of $\mathcal{T}_\mathrm{U}$.  

\begin{figure}[!htb]
\centering 
\includegraphics[scale = 0.55]{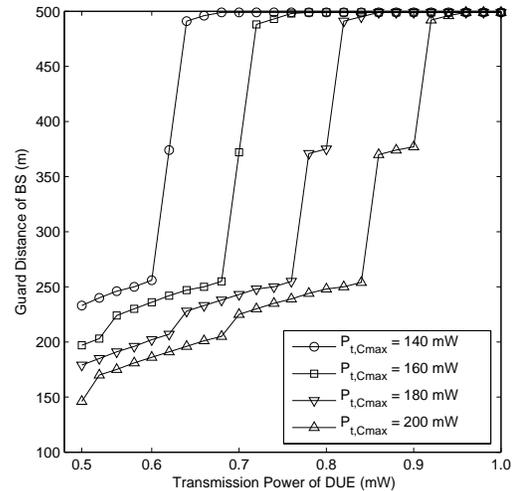} 
\caption{$P_\mathrm{t,D}$ vs. $G_\mathrm{B}$ with changed 
$P_\mathrm{t,C_{max} }$ ($R_b =2$ Mbps)} \label{fig:simu1b} 
\end{figure}

In addition to this interesting result, as we can see in Fig.
\ref{fig:simu1}, larger $P_\mathrm{t,C_{max}}$ requires a higher
$P_\mathrm{t,D}$ and offers a wider varying range to obtain the
optimal maximum throughput improvement. For example, when
$P_\mathrm{t, C_{\max}} = 140$ mW, the optimal range of
$P_\mathrm{t,D}$ is about $[0.50, 0.60]$~mW. When $P_\mathrm{t,
C_{\max}} = 200$mW, the optimal range of $P_\mathrm{t,D}$ is changed
to about $[0.70, 0.84]$ mW. Moreover, if $P_\mathrm{t,D}$ keeps
growing after exceeding its optimal range, the total throughput
improvement shrinks dramatically. This is because when all D2D pairs
are constrained near the boundary of a cell area, a slight change in
$P_\mathrm{t,D}$ will cause a striking variation on both
$G_\mathrm{B}$ and the maximum number of concurrent D2D pairs. The
relationship among $G_\mathrm{B}$, $P_\mathrm{t,D}$, and
$P_\mathrm{t,C_{max}}$ is also demonstrated in Fig. \ref{fig:simu1b}.
As stated earlier, a higher $P_\mathrm{t, C_{max}}$ leads to a shorter
$G_\mathrm{B}$, before $G_\mathrm{B}$ reaches to $r_\mathrm{C}$.
Besides, the increase of DUE transmission power leads to the increase
of $G_\mathrm{B}$, which limits the total number of D2D pairs in a
cell and the total interference at the BS. In particular,
$G_\mathrm{B}$ rises significantly after a specific value of
$P_\mathrm{t,D}$, which matches the tendency shown in Fig.
\ref{fig:simu1}.


\subsection{Impact of $R_b$ on $\mathcal{T}_\mathrm{U}$}
\begin{figure}[!htb]
\centering 
\includegraphics[scale = 0.55]{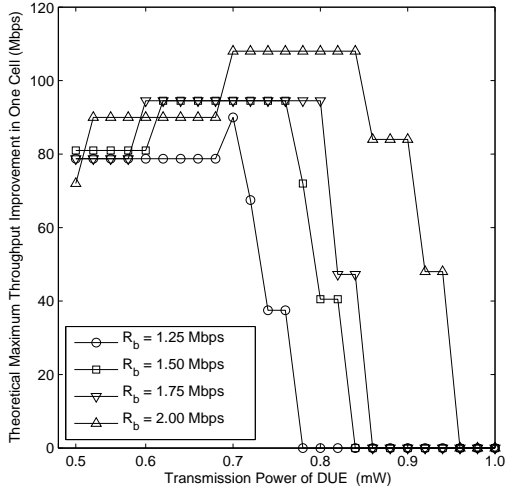} 
\caption{$P_\mathrm{t,D}$ vs. $\mathcal{T}_\mathrm{U}$ 
(when CUE is not considered) with changed $R_b$  
($P_\mathrm{t,C_{max} } =200$ mW)} \label{fig:sim2}
\end{figure}

\begin{figure}[!htb]
\centering 
\includegraphics[scale = 0.55]{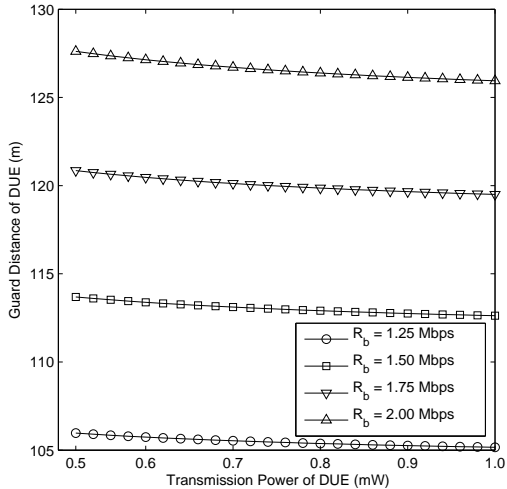} 
\caption{$P_\mathrm{t,D}$ vs. $G_\mathrm{D}$ with changed $R_b$ 
($P_\mathrm{t,C_{max} } =200$ mW)} \label{fig:sim2-rD}
\end{figure}

Similarly as the pervious evaluation group, Fig. \ref{fig:sim2} shows
the upper bound for the maximum throughput of D2D communications
varying with the changed DUE transmission power and the expected bit
rate, when the CUE's impact disk is fully included in the BS's guard
region. Generally, when $P_\mathrm{t,D}$ is fixed, a higher bit rate
$R_b$ requires a longer guard distance $G_\mathrm{D}$ between DUE nodes,
which is demonstrated in Fig. \ref{fig:sim2-rD}. The raise of $R_b$
increases the area of a D2D pair's ER, but the total throughput could
still be raised, even though the maximum number of concurrent
communicating D2D pairs in the network is reduced. Similar to the
previous group, when the DUE transmission power is monotonically
increasing for a given $R_b$, the throughput improvement grows to its
optimal first, and then drops with a substantial amount. It is worth
mentioning that the optimal maximum throughput improvement could be
obtained with a range rather than one specific value of
$P_\mathrm{t,D}$. This is due to that the number of maximum concurrent
D2D pairs does not have a continuous linear relationship with other
system parameters. Therefore, the changing of $\mathcal{T}_\mathrm{U}$
is shown as a step function. The effects of
$P_\mathrm{t,D}$ and $R_b$ on the DUE guard distance $G_\mathrm{D}$
are illustrated in Fig. \ref{fig:sim2-rD}. It is clear that
$G_\mathrm{D}$ is slowly decreased while $P_\mathrm{t,D}$ increases,
which provides stronger support for satisfying the SIR requirement of
DUEs. However, the major dominator for $G_\mathrm{D}$ is $R_b$ in our
simulation. 


\subsection{Impact of $d_\mathrm{CB}$ on General Performance Bounds}
\label{sec:simu-c}
\begin{figure}[!htb]
\centering 
\includegraphics[scale = 0.55]{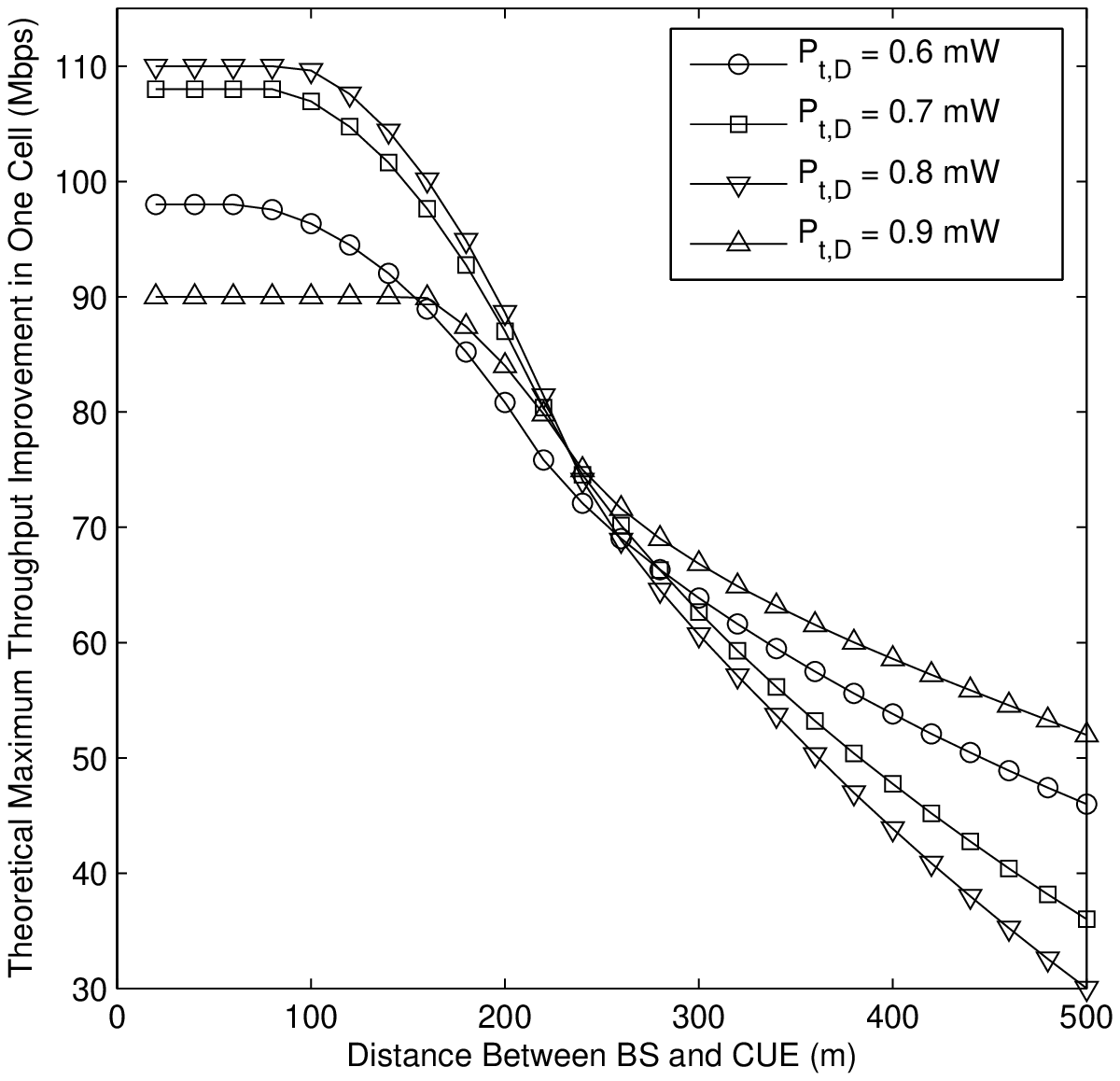} 
\caption{$d_\mathrm{CB}$ vs. $\mathcal{T}_\mathrm{U} \left(d_\mathrm{CB}, 
K\right)$ with changed $P_\mathrm{t, D}$ ($R_b = 2$ Mbps, 
$P_\mathrm{t,C_{max}} = 200$ mW)} \label{fig:sim4}
\end{figure}

In a more general scenario, the CUE could appear in any location
within its BS's coverage area. As described in Section
\ref{sec:general bounds}, since a CUE's impact disk will impair some
D2D pairs' transmissions, the maximum throughput will be affected by
$d_\mathrm{CB}$, which is the distance between BS and the
transmitting CUE. Based on (\ref{eq:upper2}), the general upper bound
of the maximum throughput improvement $\mathcal{T}_\mathrm{U}
\left(d_\mathrm{CB}, K\right)$ is calculated, and shown in Fig.
\ref{fig:sim4} to demonstrate its relationship with $d_\mathrm{CB}$
and $P_\mathrm{t, D}$. 

It is clear that the upper bound of the maximum throughput is
independent of $d_\mathrm{CB}$ at the very beginning, which
represents the scenario that the CUE's impact disk is fully included
within BS's guard region. Note that the length of the flat part of
curves in Fig. \ref{fig:sim4} is proportional to $P_\mathrm{D}$. Since
a higher $P_\mathrm{D}$ indicates a larger $G_\mathrm{B}$ as we
discussed in the previous subsection, the CUE's impact disk can
still be incorporated within the BS's guard region even with a larger
$d_\mathrm{CB}$. After that, the CUE's impact disk starts to
partially intersect with the ring area determined by $r_\mathrm{C}$
and $G_\mathrm{B}$. Thus, the maximum number of D2D pairs shrinks, and
so does the maximum throughput improvement. Compared with the results
in Fig.~\ref{fig:sim2}, the simulation results for the maximum
throughput in this figure are slightly higher, due to the different
methods of determining the maximum number of D2D pairs in
(\ref{eq:nmax}) and (\ref{eq:nmax-conti}). However, the difference is
merely the deviation about one or two D2D pairs, so it is still
acceptable. 
Similar to Fig. \ref{fig:sim2}, when the CUE's effect is excluded, the
maximum throughput improvement in a cell is developing as
$P_\mathrm{t, D}$ rises from 0.6~mW to 0.8~mW. But when $P_\mathrm{t,
D}$ is further increased to 0.9~mW, the maximum throughput falls to a
relatively low level.  

\begin{figure}[!htb]
\centering 
\includegraphics[scale = 0.55]{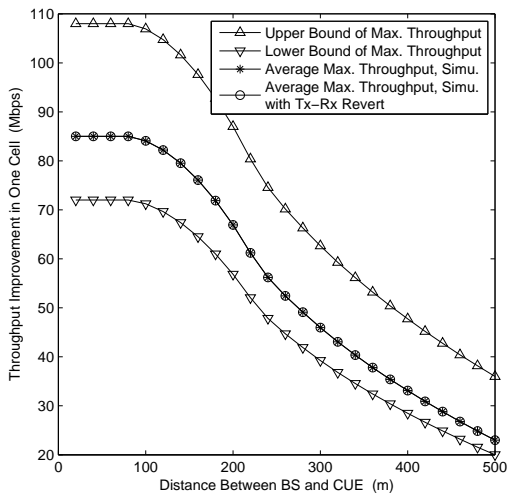} 
\caption{$d_\mathrm{CB}$ vs. Maximum throughput improvement with changed 
$d_\mathrm{D2D}$ ($R_b = 2$ Mbps, $P_\mathrm{t,C_{max}} = 200$ mW, 
$P_\mathrm{t,D} = 0.7$ mW)}\label{fig:sim5} 
\end{figure}

It should be noted that all the results above are obtained by setting
$d_\mathrm{D2D} = d_{\min}$. Therefore, the maximum number of D2D pairs
in the network reaches its upper bound. When it comes to the other
extreme scenarios i.e., $d_\mathrm{D2D} = d_{\max}$, the obtained
maximum throughput improvement turns out to be a lower bound. Both the
upper and lower bounds of the maximum throughput improvement of a
single cell are illustrated in Fig.~\ref{fig:sim5}. Moreover, the
average maximum throughput, with identical system setting but variable
$d_{\mathrm{D2D}}$ for each D2D pair, is also obtained with different
values of $d_\mathrm{CB}$. As depicted in the figure, the curve for
the average performance lies between the two bounds and closer to the
lower bound. Currently, the probability distribution of
$d_\mathrm{D2D}$ is unknown, which is usually determined by the
different application and user scenarios, so we could not provide a
closed-form expression of the average maximum throughput improvement, but this
will be addressed in our follow-on work. Besides, the 
simulation results obtained by rotating each DUE's role between 
transmitter and receiver are also shown in Fig.~\ref{fig:sim5}. It is 
clear that there is no obvious difference between the rotated and 
non-rotated results. This verifies the assumption that, the transmitter and 
receiver of a D2D pair could be treated indiscriminately in our analysis, 
which was mentioned in Section~\ref{sec:sysMod}.

\begin{figure}[!htb]
\centering 
\includegraphics[scale = 0.55]{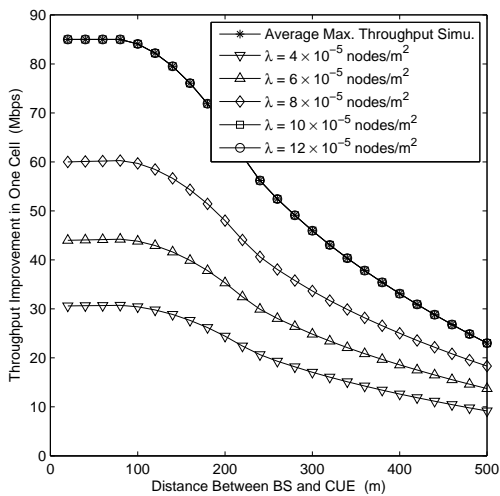} 
\caption{$d_\mathrm{CB}$ vs. Maximum throughput improvement with changed 
network density $\lambda$ ($R_b = 2$ Mbps, $P_\mathrm{t,C_{max}} = 200$ mW, 
$P_\mathrm{t,D} = 0.7$ mW)}\label{fig:sim-ppp}
\end{figure}

Considering that the number and locations of UEs in a cell may
statistically follow certain distributions, we also simulated the
throughput improvement of D2D communications in a cell when the
network nodes are distributed by following the Poisson Point Process
(PPP) with varied density~$\lambda$, which is increased from $4\times
10^{-5}$ to $12\times 10^{-5}$ nodes/m$^2$. As shown in
Fig.~\ref{fig:sim-ppp}, for different $\lambda$, the general pattern
that the total throughput improvement is decreased with increased
$d_\mathrm{CB}$ still holds. When $\lambda$ is relatively low, the
effect of $d_\mathrm{CB}$, which demonstrates the change of network
area left for D2D communications, on the throughput improvement is not
that obvious. This is because that, when all the UEs are sparsely
deployed, a D2D communication can always be successfully finished with
high probability, as long as the distance between the DUE transmitter
and receiver is within the predefined interval $[d_{\min},d_{\max}]$.
On the other hand, when the density $\lambda$ is larger than a
threshold (e.g., when $\lambda = 10\times 10^{-5}$ and $12\times
10^{-5}$ nodes/m$^2$), the number of concurrent D2D pairs in the
network (and also the throughput improvement) is then constrained by
the area possible for arranging D2D ERs, which means the changing of
node density or distribution will not affect the performance anymore.
As shown in the figure, for network scenarios with high PPP density,
the maximum throughput improvement is almost identical to the average
results illustrated in Fig.~\ref{fig:sim5}, in which all the network
nodes are uniformly distributed with a high density. Therefore, for a
given network setting, there should always be a optimal range of
network density to establish D2D communications more efficiently.

\section{Further Discussions}\label{sec:dis}

It is clear that the model and analyses mentioned above are only
focusing on a single cell with uplink resource reusing between one CUE
and multiple DUEs. However, the general method developed in this paper
could be extended to a series of more complicated but also important
scenarios. Currently, we are working on the following topics, and have
obtained some interesting results on them.

First, for more general network modeling, multiple CUEs should coexist
in the uplink transmission scenario. Due to the fact that orthogonal
channel resources are allocated to different CUEs, the DUEs reusing
different uplink resource will not interfere with each other.
Therefore, the multiple CUEs scenario is theoretically equivalent to
the combination of multiple independent single-CUE scenarios. However,
the dynamic channel allocation work for all DUEs becomes difficult,
and directly influents the final network performance improvement. We
will focus on this in the near future. 

Second, for a large network with multiple cells, which shapes a
hexagon grid, the two-hexagon-approximation used in calculating
$G_\mathrm{B}$ can be simplified to approximate the BS's guard region
only. Moreover, if each cell is further divided into several sectors
to utilize more complicated resource reusing scheme, the only change
in the analytical method is that the possible interference generated
from the neighboring cells with the same resource should be
considered. However, the symmetry feature in hexagon and the resource
reusing pattern could greatly simplify the entire analyzing process.
 
Third, if the D2D communications reuse the downlink resources rather
than the uplink ones discussed in this paper, the interfering target
is changed to the CUE(s) in the cell, and the BS's transmission may
also generate considerable interference to all the DUE receivers.
Therefore, the guard distance-based system model need to be rebuilt,
but the questions can be solved similarly by starting from the
simplified one CUE scenario, and evolved to more complicated scenarios
later.

Last but not least, although the throughput bounds do demonstrate the
extreme situations (e.g., all the ERs has the identical radius
$r_\mathrm{E,\max}$ or $r_\mathrm{E,\min}$) for a communicating D2D
pair, it will be even more useful if we could provide the probability
distribution functions (or probability mass functions) of these
performance metrics rather than some fixed values. However, for the
generalized scenario, in which each D2D pair's transmission distance
$d_\mathrm{D2D}$ is randomly selected, the total number of D2D pairs
in the finite cell or network region will be extremely difficult to be
obtained due to the packing problem and the boundary effect.
Therefore, the commonly used discrete-style interference analysis
method (i.e., obtaining each concurrent transmitter's impact on the
observed receiver individually, and adding them together), which is
also used in this paper, may not able to be applied again. This will
make the derivation for the performance metrics' distribution function
even more complicated. Interestingly, we recently found out that by
borrowing ideas from physics, the effect generated by a point
transmitter could be equalized to an area transmitter under some
conditions and vice versa \cite{Ni1406:Power}. By this means, the
accumulated interference could be obtained by an area integral, so the
complicated packing issues could be successfully rounded, and we could
have chances to obtain the desired distribution functions in a more
concise and simple way.

\section{Conclusions}\label{sec:con} 

In this paper, we have obtained the proper system settings in terms of
guard distances for BS, CUE, and DUE to ensure that multiple D2D
communications could be simultaneously carried with the traditional
uplink transmission from a CUE to the BS. In addition, we have derived
the performance bounds of the single cell's maximum throughput
improvement. Moreover, some discussions about the possible extensions
based on the current work are also given. We believe this work will
provide some useful insights for the design and optimization of more
efficient D2D communications in cellular networks.

\appendix \label{ap:intersection} 

\begin{figure}[!htb]
\centering 
\includegraphics[scale = 0.55]{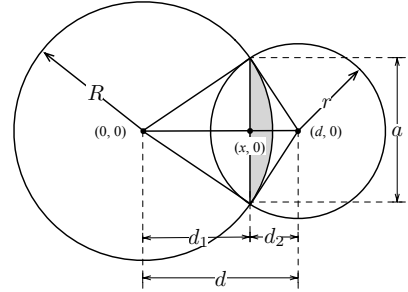} 
\caption{Circle-circle intersection} \label{fig:app_1} 
\end{figure}

As shown in Fig. \ref{fig:app_1}, two circles with radius $R$ and $r$
are centered at $(0, 0)$ and $(d, 0)$, intersecting in a lens shaped
region. The intersection points' abscissa can be calculated as
\begin{equation}
x= \frac{d^2 - r^2 + R^2}{2d}\quad,
\end{equation}
and the length of the arc in the shaded region could be calculated as
\begin{equation}
\mathcal{L}(R, r, d) = 2R\arccos\left(  \frac{d^2 - r^2 + R^2}{2dR} \right)\quad.
\end{equation}
The area of the shaded region, which is determined by $R$ and $x$, can be 
calculated as
\begin{equation}
S(R, x) = R^2 \arccos (x/R) - x \sqrt{R^2 - x^2}\quad.
\end{equation}
Similarly, the area of the other half asymmetric lens can be represented as 
$S(r, d-x)$. Therefore, the intersected area can be calculated as
\setlength{\arraycolsep}{0.0em}
\begin{eqnarray}
\mathcal{F}(R,r,d) &{}={}& S(R, x) + S(r, d-x) \nonumber\\
&{}={}& R^2 \arccos \left(\frac{d^2 + R^2 - r^2}{2dR}\right) \nonumber\\
&&{+}\: r^2 \arccos \left(\frac{d^2 + r^2 - R^2}{2dr}\right) \nonumber\\
&&{-}\:  \sqrt{4d^2 R^2 - (d^2 - r^2 + R^2)^2 } /2 \quad. \label{eq:intersect}
\end{eqnarray}
\setlength{\arraycolsep}{5pt}
For a more general case, $\mathcal{F}(R,r,d)$ can be represented as below.
\begin{equation}
\mathcal{F}(R,r,d) =  \begin{cases}
0 & \text{if~} d \geq R+r ~,\\
\text{Eq.}~(\ref{eq:intersect}), & \text{if~} R \geq r \text{~and~} R-r < d < R+r ~, \\
\pi r^2 & \text{if~} R \geq r \text{~and~} d \leq  R-r~, \\
\mathcal{F}(r,R,d) & \text{if~} R \leq r ~.
\end{cases}
\end{equation}

\end{document}